\documentclass[aps,prb,twocolumn,superscriptaddress]{revtex4-1}
\usepackage{graphicx}
\usepackage{picture}
\usepackage{xcolor}
\usepackage{color}
\usepackage{tikz}
\usepackage{amsmath}
\begin{document}

\title{Absorption Spectrum of a Two-Level System Subjected to a Periodic Pulse Sequence.}

\author{H. F. Fotso}
\affiliation{Department of Physics and Astronomy, Iowa State University, Ames, Iowa 50011, USA}
\affiliation{Department of Physics, University at Albany (SUNY),  Albany, New York 12222, USA}
\author{V. V. Dobrovitski}
\affiliation{Ames Laboratory US DOE, Ames, Iowa, 50011, USA}
\affiliation{QuTech and Kavli Institute of Nanoscience, TU Delft, Lorentzweg 1, 2628 CJ Delft, the Netherlands}

\begin{abstract}
We investigate how the quantum control of a two-level system (TLS) coupled to photons can modify and tune the TLS's photon absorption spectrum. Tuning and controlling the emission and the absorption is of much interest e.g.\ for the development of efficient interfaces between stationary and flying qubits in modern architectures for quantum computation and quantum communication. We consider the periodic pulse control, where the TLS is subjected to a periodic sequence of the near-resonant Rabi driving pulses, each pulse implementing a 180$^\circ$ rotation. For small inter-pulse delays, the absorption spectrum features a pronounced peak of stimulated emission at the pulse frequency, as well as equidistant satellite peaks with smaller spectral weights. As long as the detuning between the carrier frequency of the driving and the TLS transition frequency remains moderate, this spectral shape shows little change. Therefore, the quantum control allows shifting the absorption peak to a desired position, and locks the absorption peak to the carrier frequency of the driving pulses. Detailed description of the spectrum, and its evolution as a function time, the inter-pulse spacing and the detuning, is presented. 
\end{abstract}

\maketitle

\section{Introduction}
An interface between stationary and flying qubits, that enables a long-range entanglement between different quantum network nodes, is essential for quantum information processing \cite{JeffKimble_qtmInternet}. It is of particular importance for the solid state qubits, such as quantum dots or color centers \cite{Imamoglu_Awschalom_QDOT_PRL_99,ChildressRepeater,HansonAwschalom_QIP_ss_08,
Bernien_Hanson_Nature2013,Pfaff_Hanson_Science2014,Gao_Imamoglu_NatComm2013,Hanson_loopholeFree_Nature2015,Basset_Awschalom_PRL2011,FaraonEtalNatPhot,Sipahigil_SiV,
Rogers_SiV,SantoriVuckovicYamamoto,CarterGammonQDcavity,NV_Review_PhysRep2013}, which can be efficiently coupled to each other via photons and thus employed for quantum communications and distributed quantum information processing. However, the slow fluctuations in the environment of the solid-state qubits (e.g.\ the local strain and/or the local electric fields) constitute a lingering challenge, because they unpredictably shift the optical transition frequency of the qubits \cite{AmbroseMoerner_spectralDiffusion_Nature1991, Fu_Beausoleil_PRL2009}. This slow drift of the transition frequency (spectral diffusion) makes it difficult to achieve the precise matching between the photons originating from different qubits that is required for efficient entanglement. To mitigate the spectral diffusion problem, various methods have been proposed and successfully used \cite{Fu_Beausoleil_PRL2009,Bernien_Hanson_Nature2013,Pfaff_Hanson_Science2014,Gao_Imamoglu_NatComm2013,Hanson_loopholeFree_Nature2015,Basset_Awschalom_PRL2011,Hansom_Atature_APL2014,Acosta_Beausoleil_PRL2012, Kuhlmann_Warburton_NatPhys2013,Crooker_Bayer_PRL2010,Matthiesen_Atature_SciRep2014,FotsoEtal_PRL2016}, focusing primarily on the tuning of the emission spectrum and on improving the indistinguishability of the photons emitted from different qubits. In particular, it has been recently suggested \cite{FotsoEtal_PRL2016} that the application of a periodic sequence of the optical control pulses to a quantum emitter (a two-level system coupled to the electromagnetic radiation bath) can re-direct most of the emission into a peak located at a preset target frequency (determined by the carrier frequency of the pulse driving field), and therefore greatly improve the  indistinguishability of the photons coming from different emitters.

At the same time, there is a growing interest, accompanied by impressive progress \cite{LenhardEtal_PRA2015,TrautmannAlber_PRA2015,YangWrachtrupEtal_NatPhot2016} in the long-range entanglement schemes based on the photon absorption, and the theoretical developments which allow control and tuning of the absorption spectra have become timely and interesting. Correspondingly, a question arises whether the absorption-based entanglement can also be improved using the pulse control of the emitters, i.e. whether the absorption spectrum of a two-level system (TLS) coupled to the radiation bath can be modified and tuned by the control pulses. Besides, the studies of absorption of a TLS subjected to an external control are of fundamental interest due to the intimate connection between emission and absorption \cite{einstein_phys_Z_1917}. For instance, if the TLS is continuously driven by a strong coherent laser field then the TLS emission spectrum has an interesting three-peak structure, with two additional side peaks located at the frequencies $\pm\Omega_R$ (where $\Omega_R$ is the laser Rabi driving frequency), and the absorption spectrum of the same system also acquires additional structure, displaying the regions of gain, corresponding to an amplification of the probing weak field instead of attenuation \cite{Mollow_PhysRevA_5_1972,Wu_et_al_Mollow_PRL1977}. 

\begin{figure}
\includegraphics[angle=270,width=8.50cm]{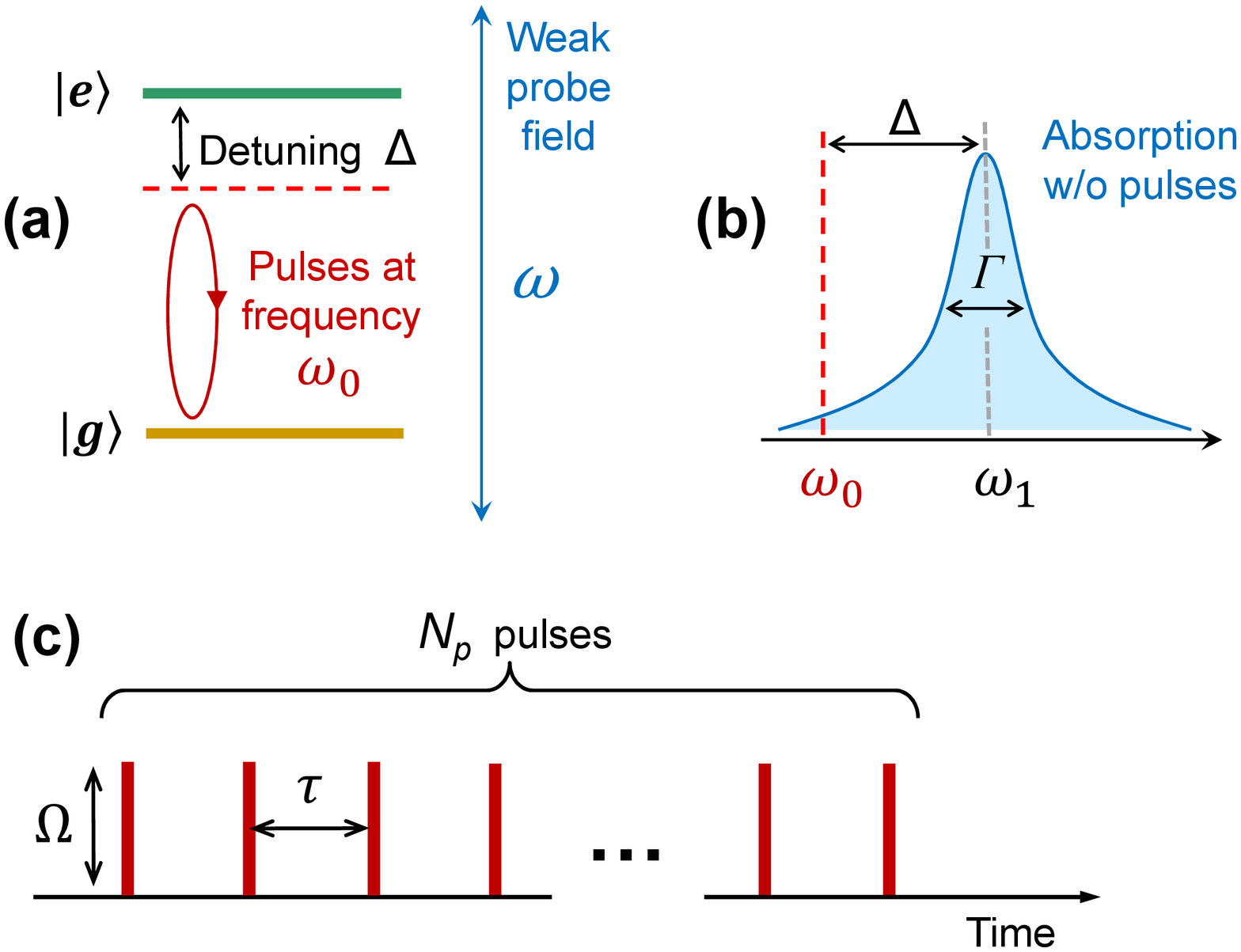}  
\caption{(Color online) (a) Schematic representation of the two-level system with ground state $|g\rangle$ and excited state $|e\rangle$ separated in the rotating frame by the detuning $\Delta$. It is probed by measuring the energy absorbed from a weak field as a function of frequency. (b) The absorption spectrum in the absence of any driving field has a Lorentzian lineshape centered around $\Delta$. (c) We will evaluate the absorption spectrum when the system is driven by a periodic sequence of $\pi$-pulses with inter-pulse time $\tau$.} 
\label{fig:skecthAbsorption}
\end{figure}

The emission spectrum of the pulse-controlled TLS exhibits similarities with the continuously driven TLS emission: it has a central peak at the carrier frequency of the pulses $\omega_0$, as well as the  satellite peaks at $\omega_0 \pm \pi/\tau, \; \pm 2\pi/\tau, \; \cdots$, where $\tau$ is the inter-pulse distance. Thus, it is reasonable to expect that absorption also can be controlled with the periodic pulses, and that the resulting absorption spectrum also has non-trivial features. In this work we study the absorption spectrum of a TLS driven by a periodic sequence of optical $\pi$-pulses, and examine its dependence on the pulse sequence period and the detuning of the emitter with respect to the pulse frequency (Fig.\ref{fig:skecthAbsorption}). We show that both expectations above are correct, and therefore the pulse control indeed can be a useful tool for controlling and tuning the absorption spectrum of a TLS. The absorption spectrum has a pronounced peak of stimulated emission at the carrier frequency of the pulses, and equidistant satellite peaks with smaller spectral weights. The qualitative features of this absorption spectrum do not change much as long as the detuning between the carrier frequency of the driving pulses and the TLS transition frequency remains moderate. Therefore, we show that the optical control enables creation of pairs of quantum nodes (one node working as an emitter and the other as an absorber) with precisely matching frequencies, and therefore greatly increased entanglement efficiency. This approach can also be used to improve the coupling of the emitters and the absorbers to the optical cavities, since the laser pulses can tune both the emission and the absorption lines of the respective quantum nodes, bringing them in the resonance with the respective cavities, and stabilizing both the emission and the absorption peaks at the desired location.

The rest of the paper is organized as follows. In Sec.~\ref{sec:Model} we describe the model of the two-level system coupled to the photon bath and controlled by the pulses, the master equations governing the system dynamics, and the two methods, analytical and numerical, used for calculating the absorption spectrum. In Sec.~\ref{sec:Results} we present analytical and numerical results demonstrating the control and tunability of the absorption spectrum. In Sec.~\ref{sec:Conclusions} we present conclusions.

\section{Model of the two-level system coupled to the electromagnetic radiation bath}
\label{sec:Model}

We model the quantum emitter as a TLS with the ground state $|g\rangle$ and the excited state $|e\rangle$, separated in energy by $E_e - E_g = \hbar\omega_1$; below we set $\hbar=1$. Initially, at time $t=0$, the excited state is occupied and the ground state is empty. The TLS is coupled to a photon bath, and is periodically driven by pulses of the laser field with the Rabi frequency $\Omega$. Within the rotating-wave approximation (RWA) \cite{Cohen_Tannoudji_Book1992}, in the reference frame rotating at frequency $\omega_0$, the system in question is described by the Hamiltonian 
\begin{eqnarray}
\label{eq:hamiltonian}
H = \sum_{k} \omega_k a^{\dagger}_{k}a_{k} &+& \frac{\Delta}{2} \sigma_z - i \sum_{k} g_{k} \left( a^{\dagger}_{k} \sigma_- - a_{k} \sigma_+ \right) \nonumber \\
&+& \frac{\Omega_x(t)}{2}(\sigma_+ + \sigma_-), 
\end{eqnarray}
where $\Delta=\omega_1-\omega_0$ is the detuning of the TLS's transition frequency from the carrier frequency of the pulses; here we introduced the standard pseudo-spin Pauli operators for the TLS, namely $\sigma_z = |e\rangle \langle e| - |g\rangle \langle g|$, $\sigma_+ = |e\rangle \langle g|$ and $\sigma_- = |g\rangle \langle e| = (\sigma_+)^{\dagger} $. Furthermore, $a^{\dagger}_{k}$ and $a_{k}$ are respectively the creation and the annihilation operator for a photon of mode $k$ with the frequency $\omega_k$, and $g_{k}$ is the strength of coupling to the TLS. Note that in the rotating frame all frequencies are measured from the pulse carrier frequency $\omega_0$, so that the zero frequency in the rotating frame corresponds to $\omega_0$ in the lab frame; we take it as the target frequency for our TLS. 

The time-dependent driving $\Omega_x(t)$ in Eq.~(\ref{eq:hamiltonian}) represents the control pulses; here we consider the simple situation of the square-shaped pulses, with $\Omega_x(t)=\Omega$ during the pulses and zero otherwise. In fact, below we assume that the pulses are almost instantaneous, i.e.\ that $\Omega$ is much larger than all other relevant energy scales, and that each pulse performs an almost instantaneous 180$^\circ$ rotation of the TLS around the $x$-axis, interchanging $|e\rangle$ and $|g\rangle$; this assumption will be discussed further below. In the absence of control ($\Omega_x(t)\equiv 0$), the system exhibits spontaneous decay, and the corresponding emission rate is $\Gamma = 2\pi\int g_k^2 \; \delta(\omega_k - \Delta) \; dk$; we normalize our energy and time units so that $\Gamma = 2$, and the corresponding spontaneous emission line has a simple Lorentzian shape $1/(\omega^2 + 1)$, with the half-width equal to 1. The absorption spectrum is defined here in a standard way, as the energy absorbed by the TLS from a weak probing field of frequency $\omega$. The probing field is assumed to be weak enough that it does not significantly affect the population of each state~\cite{Mollow_PhysRevA_3_1972, Mollow_PhysRevA_5_1972}; our goal is to calculate the absorption as a function of frequency and time.

To understand the dynamics of the system, we analyze the time evolution of the density matrix of the emitter, which is written as
\begin{eqnarray}
 \rho(t) &=& \rho_{ee}(t) |e\rangle \langle e| + \rho_{eg}(t) |e\rangle \langle g| \nonumber \\
 &+& \rho_{ge}(t) |g \rangle \langle e|  + \rho_{gg}(t) |g\rangle \langle g| \; ,
\label{eq:TrueDensMatr}
\end{eqnarray}
with $\rho_{ge}^* = \rho_{eg}$. For the TLS described by the above Hamiltonian (\ref{eq:hamiltonian}), within the Markovian approximation, the density matrix operator is governed by the master  equations~\cite{Cohen_Tannoudji_Book1992} in the rotating-wave approximation:
\begin{equation}
\label{eq:MasterEquation_1} 
\begin{split}
\dot{\rho}_{ee} &= i \frac{\Omega_x(t)}{2}(\rho_{eg} - \rho_{ge}) - \Gamma \rho_{ee} \; ,   \\
\dot{\rho}_{gg} &= -i \frac{\Omega_x(t)}{2}(\rho_{eg} - \rho_{ge}) + \Gamma \rho_{ee} \; ,  \\
\dot{\rho}_{ge} &= (i\Delta -\frac{\Gamma}{2}) \rho_{ge} -i \frac{\Omega_x(t)}{2}\left(\rho_{ee} - \rho_{gg}\right) \; , \\
\dot{\rho}_{eg} &= (-i\Delta -\frac{\Gamma}{2}) \rho_{eg} +i \frac{\Omega_x(t)}{2}\left(\rho_{ee} - \rho_{gg}\right) .
\end{split}
\end{equation}
Since the pulse driving is assumed to be strong and short ($\Omega \gg \Delta, \Gamma$), the pulses can be considered as instantaneous. Each of them inverts the populations of the excited and ground state and swaps the values of $\rho_{eg}$ and $\rho_{ge}$, i.e. 
\begin{equation}
 \rho(n\tau + 0) = \sigma_x \rho(n\tau - 0) \sigma_x
 \label{eq:pulseEffect}
\end{equation}
where $\rho(n\tau - 0)$ and $\rho(n\tau + 0)$ are the density matrices immediately before and after the pulse, correspondingly, with $n$ being an integer and $\tau$ the period of the pulse sequence; in other words, the pulses interchange $\rho_{ee}$ with $\rho_{gg}$, and $\rho_{eg}$ with $\rho_{ge}$. 

We want to determine the energy absorbed from a weak probing field by the TLS subjected to the periodic sequence of the $\pi$-pulses. Since the effect of the probing field is small, the absorption spectrum can be calculated within the linear response theory, so that at long times $T$ the absorbed energy $Q(\omega)$ is given by~\cite{Mollow_PhysRevA_3_1972,Mollow_PhysRevA_5_1972}
\begin{eqnarray}
Q(\omega) &=& 2 A^2 \\
&\times & \mathrm{Re} \left\{ \int_0^T  dt \int_0^{T-t} d\theta \; \langle \left[ \sigma_-(t) , \sigma_+(t+\theta) \right] \rangle \mathrm{e}^{-i \omega \theta} \right\}, \nonumber
\label{eq:absorptionEq}
\end{eqnarray}
where $[O_1, O_2]$ is the commutator of the operators $O_1$ and $O_2$, and the angled brackets represent the expectation values evaluated in the absence of the probing field. $\sigma_-(t)$ and  $\sigma_+(t+\theta)$ are the time-dependent operators in the Heisenberg representation, and the expectation values are taken with respect to the initial state of the two-level system (in our case, fully occupied excited state and empty ground state). The constant $A$ is independent of the pulse parameters, and does not affect the spectral shape, determining only the absolute scale of the absorption. The expression (\ref{eq:absorptionEq}) can be rewritten as
\begin{eqnarray}
Q(\omega) &=& 2 A^2 \mathrm{Re} \left\{ \mathcal{P}_2(\omega) - \mathcal{P}_1(\omega) \right\}  \nonumber \\
            &=& P_2(\omega) - P_1(\omega)
\label{eq:absorptionEq2}
\end{eqnarray}
where
\begin{equation}
 \mathcal{P}_2(\omega) =  \int_0^T \; dt\; \int_0^{T-t} d\theta \; \langle  \sigma_-(t) \sigma_+(t+\theta) \rangle \mathrm{e}^{-i \omega \theta}
\end{equation}
and
\begin{equation}
 \mathcal{P}_1(\omega) =  \int_0^T \; dt\; \int_0^{T-t} d\theta \; \langle \sigma_+(t+\theta) \sigma_-(t) \rangle \mathrm{e}^{-i \omega \theta}
\end{equation}
The term $P_1(\omega)=2 A^2 \mathrm{Re} \left\{ \mathcal{P}_1(\omega) \right\} $ can be viewed as the direct emission of the two-level system and $P_2(\omega)=2 A^2 \mathrm{Re} \left\{ \mathcal{P}_2(\omega) \right\}$ as the direct absorption so that the difference yields the net absorption~\cite{RF_Mollow_PhysRev1969}. We evaluate the terms $P_1(\omega)$ and $P_2(\omega)$ separately, and obtain the total absorption spectrum $Q(\omega)$ by taking the difference.

\begin{figure}
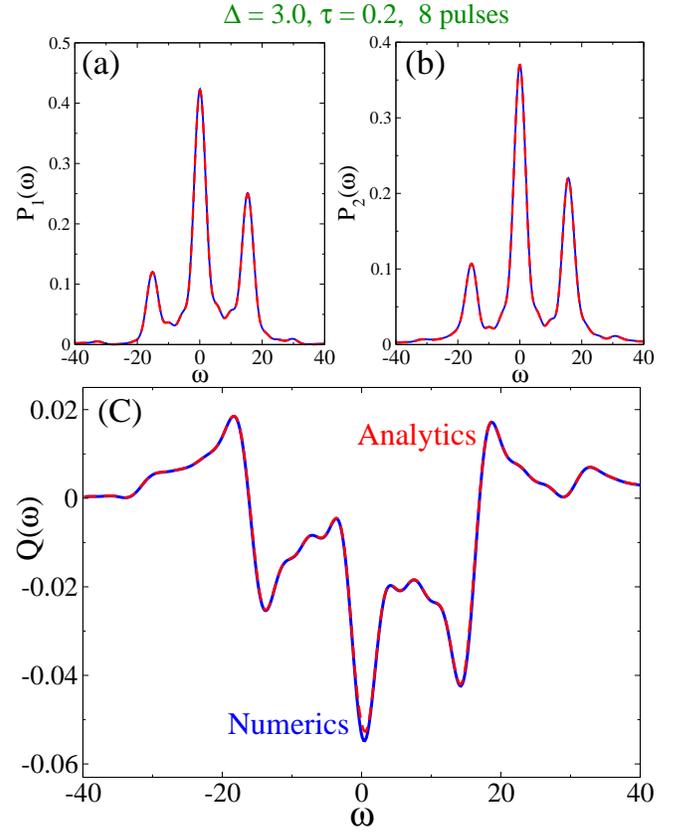

\includegraphics*[width=8.5cm]{figure2_a_b.eps}
\includegraphics*[width=8.5cm]{figure2_c.eps}
\caption{(Color online) Absorption spectrum of a two-level system with detuning $\Delta = 3.0$ driven by a periodic sequence of $\pi$-pulses of period $\tau = 0.2$ after $N_p=8$ pulses. Panels (a) and (b) show the terms $P_1(\omega)$ (direct emission) and $P_2(\omega)$ (direct absorption), correspondingly, and panel (c) shows the difference between the two terms, which is the total absorption $Q(\omega)$. The results are obtained by solving the master equation numerically (blue) and analytically in the limit of a large number of pulses (red dashed). The two approaches give very close results despite the assumed limit $N_p\gg 1$ in the analytical result and finite timestep used in the Fourier transform of the numerical results.} 
\label{fig:numerics_analytics}
\end{figure}

To evaluate the emission spectrum, it is convenient to re-express the two-time correlation function $\langle \sigma_+(t+\theta) \sigma_-(t) \rangle$ as a single-time expectation value~\cite{RF_Mollow_PhysRev1969,Scully_Zubairy_book1997,Loudon_book1983}, according to
\begin{eqnarray}
&\ &\langle\sigma_+(t+\theta) \sigma_-(t) \rangle = \nonumber \\
&\ &\ \ =\mathrm{Tr} \left[ \rho(0) U^{-1}(0,t+\theta) \sigma_+ U(0,t+\theta)U^{-1}(0,t) \sigma_- U(0,t)\right] \nonumber \\
&\ &\ \ = \mathrm{Tr} \left[ \sigma_- \rho(t) U^{-1}(t,t+\theta) \sigma_+ U(t,t+\theta) \right] \nonumber \\ 
&\ &\ \ = \mathrm{Tr} \left[ \rho'(t,t + \theta) \sigma_+ \right] \label{eq:p_sigmaP_sigmaM} 
\end{eqnarray}
where $\sigma_+$ and $\sigma_-$ are the time-independent Pauli operators in the Schr\"odinger representation, and $U(t_1,t_2)$ is the evolution operator of the emitter from time $t_1$ to time $t_2$, as  determined by the master equations  (\ref{eq:MasterEquation_1}). The calculations are simplified by introducing the matrix $\rho'(t,s)$; its initial value at $s=t$ is $\rho'(t,t)=\sigma_-\rho(t)$, and its further evolution from $s=t$ to $s=t+\theta$ is governed by the emitter's evolution operator $U(t,t+\theta)$, so that $\rho'(t,t+\theta)=U(t,t+\theta)\rho'(t,t) U^{-1}(t,t+\theta)$. In this way the evaluation of the two-time correlators becomes rather straightforward (although lengthy, see Appendix for details), and the function $P_1(\omega)$ can be obtained analytically and/or numerically. In order to calculate the function $P_2(\omega)$, we use the same procedure, simplifying the two-time correlation function as
\begin{equation}
 \langle \sigma_-(t) \sigma_+(t+\theta) \rangle = \mathrm{Tr} \left[ \sigma_+ \rho''(t,t + \theta) \right],
 \label{eq:p_prime_sigmaM_sigmaP}
\end{equation}
by introducing the matrix $\rho''(t) = \rho(t)\sigma_-$, whose time evolution is also governed by $U(t,t+\theta)$, i.e.\ $\rho''(t,t+\theta)=U(t,t+\theta) \rho''(t,t) U^{-1}(t,t+\theta)$. Note that $\rho'$ and $\rho''$ are not density matrices, and the symmetries of the proper density matrix $\rho(t)$ (such as $\rho_{gg}=1-\rho_{ee}$ and/or $\rho_{ge}^* =\rho_{eg}$) are not applicable to $\rho'$ and $\rho''$.

\begin{figure}
\includegraphics[width=8.50cm]{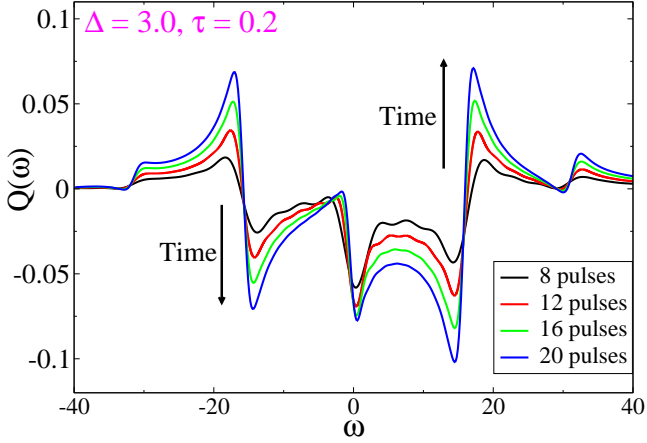}  
\caption{(Color online) Absorption spectrum of a two-level system with detuning $\Delta = 3.0$, driven by a periodic sequence of $\pi$-pulses with period $\tau = 0.2$, after $N_p=8$  (black), $N_p=12$ (red), $N_p=16$ (green), and $N_p=20$ (blue) pulses. The curves present analytical results, the arrows indicate the increasing number of pulses (increasing total time of the sequence).} 
\label{fig:Delta3p0Tau0p2_4XNp}
\end{figure}

In the absence of the pulses, the absorption spectrum has a Lorentzian-shaped profile centered at the emitter's frequency that equals to the detuning $\Delta$ (Fig.\ref{fig:skecthAbsorption}). In the presence of the pulses, we calculated the absorption spectrum both analytically and numerically by iteratively evolving the density matrix operator between successive pulses on a discrete time grid, using the equations of motion (\ref{eq:MasterEquation_1}), with the initial conditions $\rho_{ee} = 1$, $\rho_{eg}=\rho_{ge}=\rho_{gg}=0$, and then making use of (\ref{eq:p_sigmaP_sigmaM}) and (\ref{eq:p_prime_sigmaM_sigmaP}) to calculate the two-time correlation functions.\\

\subsection{Numerical solution}

To find the solution numerically, we divide the time axis in the intervals of length $\tau$ (equal to the inter-pulse separation), and each interval between the pulses is further discretized into smaller steps of length $\Delta t$. The goal is to find the two-time correlators $\langle \sigma_-(t) \sigma_+(t+\theta) \rangle$ and $\langle \sigma_-(t) \sigma_+(t+\theta) \rangle$ for each value of $t$ and $\theta$ on this time grid, and use Fourier transform to find $P_1(\omega)$ and $P_2(\omega)$, whose difference gives the absorption spectrum $Q(\omega)$. 

We start at $t=0$ with the known initial conditions for $\rho(t)$, and use Eqs.~\ref{eq:MasterEquation_1} to evolve all elements of the density matrix $\rho(t)$ from time $t$ to $t+\Delta t$, and repeat this integration up to $t = \tau$. Then the $\pi$-pulse is applied to the system, transforming the density matrix in accordance with Eq.~(\ref{eq:pulseEffect}), and the iterative integration is resumed to propagate the density matrix from $t=\tau$ to $t=2\tau$, until another pulse is applied at $2\tau$. The process is repeated until time $T = N_p\tau$ is reached, where $N_p$ is the total number of pulses. 
In this way we can obtain the elements of $\rho'(t,t)$ and $\rho''(t,t)$ for every $t\in[0,N_p\tau]$. Then, for each time $t$ we propagate the matrices $\rho'$ and $\rho''$ from time $t$ to time $t+\theta$ by solving the master equations (\ref{eq:MasterEquation_1}); the values $\rho'(t,t)$ and $\rho''(t,t)$ serve as initial conditions. As a result, we obtain $\rho'(t,t+\theta)$ and $\rho''(t,t+\theta)$ for all values of $\theta\in [0,T-t]$. This procedure produces the two-time correlators $\langle\sigma_+(t+\theta) \sigma_-(t)\rangle$ and $\langle\sigma_-(t) \sigma_+(t+\theta)\rangle$, see Eqs.~(\ref{eq:p_sigmaP_sigmaM}) and (\ref{eq:p_prime_sigmaM_sigmaP}). Finally, Fourier transform with respect to $\theta$ and integration over $t$ give us $\mathcal{P}_1(\omega)$ and $\mathcal{P}_2(\omega)$, thus determining the absorption spectrum $Q(\omega)$.

\subsection{Analytical solution}

The analytical solution for the density matrix evolution between the pulses can be obtained directly from Eqs.~\ref{eq:MasterEquation_1}, and combined with the analytically calculated transformation of the density matrix by pulses as described by Eq.~(\ref{eq:pulseEffect}), thus providing a fully analytical solution to the problem. The corresponding calculation is quite lengthy, and is presented in detail in the Appendix. In the limit of long $T$ (i.e.\ large number of pulses $N_p$), the resulting expression for $\mathcal{P}_1(\omega)$ is
\begin{widetext}
\begin{eqnarray}
 \mathcal{P}_1(\omega) &=& \frac{1}{(1 + \mathrm{e}^{-\Gamma \tau})\gamma_0} \Bigg[ \left( \frac{1-\mathrm{e}^{-\Gamma \tau}}{\Gamma} - \mathrm{e}^{-\gamma_0 \tau} \frac{\mathrm{e}^{\gamma_2 \tau}-1}{\gamma_2}
 + \frac{\mathrm{e}^{\gamma_2 \tau}-1}{\gamma_2}\,\frac{1-\mathrm{e}^{-\gamma_0 \tau}}{\mathrm{e}^{2 \gamma_1 \tau}-1} \right) 
\left(N_p + \frac{\mathrm{e}^{-\Gamma \tau}}{1 + \mathrm{e}^{-\Gamma \tau}} \right) \nonumber \\
&-& \frac{\mathrm{e}^{\gamma_2 \tau}-1}{\gamma_2}\, \frac{1-\mathrm{e}^{-\gamma_0 \tau}}{\mathrm{e}^{2 \gamma_1 \tau}-1} \, \left(  
2\,\frac{\mathrm{e}^{-N_p \gamma_1 \tau} - 1 }{ \mathrm{e}^{-2 \gamma_1 \tau } - 1} + (\mathrm{e}^{-\Gamma \tau} - \mathrm{e}^{ -2 \Gamma \tau})\frac{ \mathrm{e}^{-N_p \gamma_1 \tau } }{\mathrm{e}^{-2 \gamma_1 \tau} - \mathrm{e}^{-2 \Gamma \tau } } \right) 
\Bigg]
\label{eq:P_omega} 
\end{eqnarray}
We have also performed the similar calculation for $\mathcal{P}_2(\omega)$, expressing it as  $\mathcal{P}_2(\omega)=\mathcal{P}_3(\omega)-\mathcal{P}_1(\omega)$, and the resulting expression for $\mathcal{P}_3(\omega)$ in the long-time limit is
\begin{eqnarray}
 \mathcal{P}_3 =  \frac{N_p\tau}{\gamma_0} -  \frac{N_p}{\gamma_0^2}(1 - \mathrm{e}^{-\gamma_0 \tau })
+ \frac{\mathrm{e}^{\gamma_0 \tau}+\mathrm{e}^{-\gamma_0 \tau}-2}{\gamma_0^2(\mathrm{e}^{2\gamma_1\tau}-1)}\, \left[N_p - \frac{2}{1-\mathrm{e}^{-2\gamma_1\tau}}\right]
\label{eq:P_prime_omega2}
\end{eqnarray}
\end{widetext}
where $\gamma_0 = i(\omega -\Delta ) + \Gamma/2$, $\gamma_1 = i\omega  + \Gamma/2$, and $\gamma_2 = i(\omega -\Delta ) - \Gamma/2$. 

\section{Results}
\label{sec:Results}

Fig.~\ref{fig:numerics_analytics} shows the absorption spectrum obtained using both analytical and numerical approaches for a two-level system with $\Delta = 3.0$ and a pulse sequence with $\tau = 0.2$ after $N_p=8$ pulses. The panels \ref{fig:numerics_analytics}(a) and \ref{fig:numerics_analytics}(b) show $P_1(\omega)$ and $P_2(\omega)$, respectively, while the panel \ref{fig:numerics_analytics}(c) shows the absorption spectrum obtained by taking their difference according to Eq.~\ref{eq:absorptionEq2}. In the presence of the pulse control we see the main peak in the absorption spectrum at the carrier frequency of the pulses ($\omega = 0$ in the rotating frame), and the satellite peaks at the multiples of $\pm \pi/\tau$. A good agreement is clearly seen, despite the large number of pulse limit used in the analytical result and the numerical Fourier transform of the finite time step data in the numerical result. The agreement is further improved by considering the spectrum at longer times, see Appendix. These results provide clear validation of the tools used in these studies. Moreover, note that the absorption spectrum, as given by Eq.(\ref{eq:absorptionEq}), is the difference of two terms of comparable magnitude in a broad frequency range. As a result, it is influenced by numerical errors, but the small discrepancy between the analytical and the numerical results shows that this kind of errors is not critical. Thus, the numerical solution can be used in the future studies of more complex driving protocols, which may not be amenable to an analytical solution. 

\begin{figure}
\includegraphics[width=8.50cm]{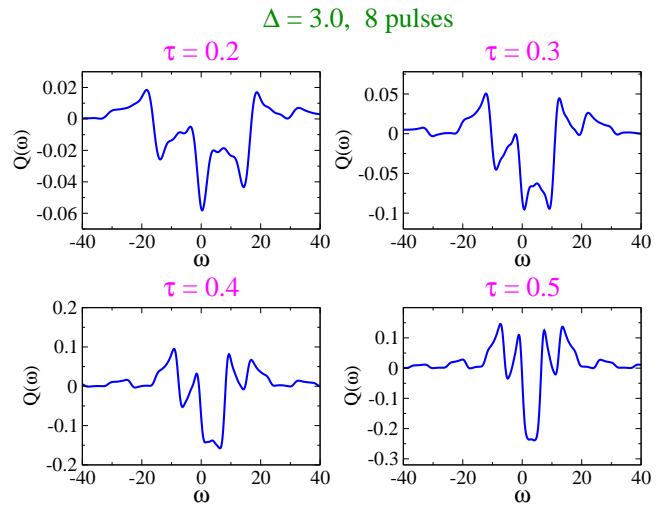}  
\caption{(Color online) Absorption spectrum of a two-level system with detuning $\Delta = 3.0$ driven by a periodic sequence of $\pi$-pulses with the periods $\tau = 0.2$, $\tau = 0.3 $, $\tau = 0.4$, and $\tau = 0.5 $, after $N_p=8$ pulses. The curves present analytical results.} 
\label{fig:Delta3p0Np8_4xTau}
\end{figure} 

Fig.\ref{fig:Delta3p0Tau0p2_4XNp} shows the analytic results for the time evolution of the absorption spectrum for $\Delta = 3.0$ and $\tau = 0.2$. The snapshots of the spectrum are presented after $N_p=8$  (black), $N_p=12$ (red), $N_p=16$ (green), and $N_p=20$ (blue) pulses. The absorption spectra feature a positive part and a negative part. The latter corresponds to the stimulated emission and the former to the ''true absorption''~\cite{BloomMargenau_PhysRev1953, Mollow_PhysRevA_5_1972}. The central peak corresponds to the stimulated emission at the pulse frequency and satellite peaks at multiples of $\pm \pi/\tau$ with amplitudes that decrease away from the central frequency and are strongly suppressed at large frequencies. The lineshape is established early, and the amplitude of the peaks increases with time.

\begin{figure}
\includegraphics[width=8.50cm]{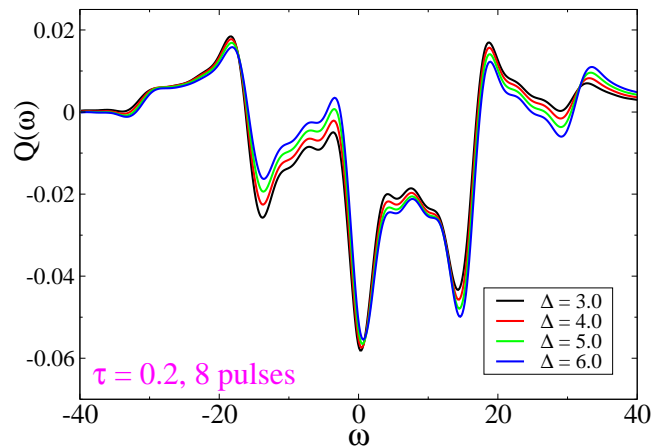}  
\caption{(Color online) Absorption spectrum of the TLS with detunings $\Delta = 3.0 $(black), $\Delta = 4.0 $(red), $\Delta = 5.0 $(green), and $\Delta = 6.0 $(blue) driven by a periodic sequence of $\pi$-pulses of period $\tau = 0.2$ after $N_p=8$ pulses. The curves present analytical results.} 
\label{fig:Tau0p20Np8_4xDelta}
\end{figure}

In Fig.\ref{fig:Delta3p0Np8_4xTau} we present the dependence of the absorption spectrum for $\Delta = 3.0$ on the period $\tau$ of the pulse sequence. The absorption spectrum is shown after 8 pulses for $\tau = 0.2$, $\tau = 0.3 $, $\tau = 0.4$, and $\tau = 0.5$. The satellite peaks move closer to the central peak and their relative amplitude increases as $\tau$ becomes longer. Also note the increase in the positive fraction of spectral weight with increasing $\tau$.

It is also interesting to study the dependence of the absorption spectrum on the detuning $\Delta$. The corresponding results are shown in Fig.\ref{fig:Tau0p20Np8_4xDelta} which presents the spectrum under a pulse sequence of period $\tau = 0.2$ for the detuning values of $\Delta = 3.0 $(black), $\Delta = 4.0 $(red), $\Delta = 5.0 $(green), and $\Delta = 6.0 $(blue) after 8 pulses. The lineshape remains almost the same for all four values of the detuning parameter when $\tau$ is kept constant. In fact, we observe that the lineshape of the absorption spectrum shows little dependence on $\Delta$ as long as $\Delta \cdot \tau \lesssim 1$.  The figure shows that the fraction of the spectral weight contained in the positive-frequency satellites (with $\omega>\omega_0$) slightly increases with $\Delta$, while the spectral weight of the negative-frequency satellites ($\omega<\omega_0$) correspondingly decreases.

\section{Conclusions}
\label{sec:Conclusions}

We have studied the absorption spectrum of a two-level system driven by a periodic sequence of $\pi$-pulses. This absorption spectrum is determined by the energy absorbed from a probing field weak enough to not significantly affect the population of the excited and ground state. We have solved the problem by integrating the master equation analytically and numerically and obtained from both methods results that are in excellent agreement. Our results show that for moderate values of $\Delta \cdot \tau$, the absorption spectrum has a lineshape with little dependence on $\Delta$. It has a pronounced peak of stimulated emission at a the pulse frequency along with satellite peaks at multiples of $\pm \pi/\tau$ away from this frequency. The weights of these satellite peaks are strongly suppressed away from the central peak. 
By using the optical control considered in this work (with, possibly, more complex pulse protocols), it is possible to create pairs of quantum nodes, with one node working as an emitter and the other as an absorber, with precisely matching frequencies, and therefore greatly increased entanglement efficiency. In a similar manner, one can use it to improve the coupling of the emitters and absorbers to the optical cavities, using the laser pulses to tune both the emission and the absorption lines of the respective quantum nodes, bringing them in the resonance with the respective cavities, and stabilizing both the emission and the absorption peaks at the desired location.

\section*{Acknowledgments} 

We thank D. D. Awschalom and M. E. Flatt\'e for helpful discussions. Work at the Ames Laboratory was supported by the US Department of Energy, Office of Science, Basic Energy Sciences, Division of Materials Sciences and Engineering. The Ames Laboratory is operated for the US Department of Energy by Iowa State University under Contract No. DE-AC02-07CH11358. This work was partially supported by AFOSR MURI program.

\appendix
\section*{Appendix}
Here we present the details of the analytical calculation of the absorption spectrum for a two-level system subjected to a periodic sequence of control pulses. As shown in Refs.~\onlinecite{Mollow_PhysRevA_5_1972,BloomMargenau_PhysRev1953}, the absorption spectrum can be determined from the two-time correlation functions of the TLS:
\begin{eqnarray}
Q(\omega) &=& 2A^2 \times \\ \nonumber
&\ &\mathrm{Re} \left\{ \int_0^T  dt \int_0^{T-t} d\theta \; \langle \left[ \sigma_-(t) , \sigma_+(t+\theta) \right] \rangle \mathrm{e}^{-i \omega \theta} \right\},
 \label{eq:absorptionEquationAppendix}
\end{eqnarray}
where $[ , ]$ is the commutator of the two enclosed operators, and the angled brackets represent the expectation values evaluated in the absence of the probing field. $A$ is a proportionality constant. This can be rewritten as:
\begin{eqnarray}
  Q(\omega) &=& 2 A^2 \mathrm{Re} \left\{ \mathcal{P}_2(\omega) - \mathcal{P}_1(\omega) \right\}  \label{eq:absorptionEq2Appendix}  \\
            &=& P_2(\omega) - P_1(\omega),  \nonumber
\end{eqnarray}
with
\begin{equation}
 \mathcal{P}_2(\omega) =  \int_0^T \; dt\; \int_0^{T-t} d\theta \; \langle  \sigma_-(t) \sigma_+(t+\theta) \rangle \mathrm{e}^{-i \omega \theta}
\end{equation}
and
\begin{equation}
 \mathcal{P}_1(\omega) =  \int_0^T \; dt\; \int_0^{T-t} d\theta \; \langle \sigma_+(t+\theta) \sigma_-(t) \rangle \mathrm{e}^{-i \omega \theta}.
\end{equation}
The terms $P_1(\omega)=2 A^2 \mathrm{Re} \left\{\mathcal{P}_1(\omega) \right\} $ and $P_2(\omega)=2 A^2 \mathrm{Re} \left\{\mathcal{P}_2(\omega) \right\} $ can be evaluated separately and the absorption spectrum obtained by taking the difference. To find $\mathcal{P}_2(\omega)$, we express the correlation function as
\begin{widetext}
\begin{eqnarray}
 \langle \sigma_-(t) \sigma_+(t+\theta) \rangle &=& \mathrm{Tr} \left[ \rho(0) U^{-1}(0, t) \sigma_- U(0, t) U^{-1}(0, t+\theta) \sigma_+ U(0, t+\theta)\right] \\ \nonumber
&=& \mathrm{Tr} \left[ \sigma_+  U(t, t+\theta) U(0, t) \rho(0) U^{-1}(0, t) \sigma_- U(0, t)U^{-1}(0, t) U^{\dagger}(t, t+\theta) \right] \\ \nonumber
&=& \mathrm{Tr} \left[ \sigma_+ U(t, t+\theta) \rho(t)\sigma_-  U^{-1}(t, t+\theta)\right] \\ \nonumber
&=& \mathrm{Tr} \left[ \sigma_+ U(t, t+\theta) \rho''(t,t) U^{-1}(t, t+\theta)\right] \\ \nonumber
&=&\mathrm{Tr} \left[ \sigma_+ \rho''(t,t+\theta)\right] 
\end{eqnarray}
\end{widetext}
where $\sigma_+$ and $\sigma_-$ are the Pauli operators, and $U(t_1,t_2)$ is the operator of the emitter's evolution from $t_1$ to $t_2$, as determined by the master equations (\ref{eq:MasterEquation_1}). The subsequent calculations are facilitated by introducing the matrix $\rho''(t,s)$; its initial value at $s=t$ is defined as $\rho''(t,t)=\rho(t)\sigma_-$, and its further evolution from $s=t$ to $s=t+\theta$ is governed by the emitter's evolution operator $U(t,t+\theta)$, so that $\rho''(t,t+\theta)=U(t,t+\theta)\rho''(t,t) U^{-1}(t,t+\theta)$. 

It is informative to write $\rho''(t,s)$ explicitly as
\begin{equation}
 \rho''(t,s) = \begin{pmatrix} \rho''_{ee}(t,s) & \rho''_{eg}(t,s) \\ \rho''_{ge}(t,s) & \rho''_{gg}(t,s) \end{pmatrix}
\end{equation}
so that
\begin{equation}
 \sigma_+ \rho^{\prime\prime}(t + \theta) = \begin{pmatrix} \rho''_{ge}(t,t+ \theta) & \rho^{\prime\prime}_{gg}(t,t+ \theta) \\ 0 & 0 \end{pmatrix},
\end{equation}
and the corresponding two-time correlation function is obtained directly as
\begin{eqnarray}
\langle\sigma_-(t)\sigma_+(t+\theta)\rangle &=& \mathrm{Tr} \left[ \sigma_+ \rho''(t,t+\theta)\right] \nonumber \\
&=& \rho''_{ge}(t,t+ \theta).
\end{eqnarray}
The initial condition for $\rho''$, corresponding to $s=t$, has a form
\begin{equation}
 \rho''(t,t) = \rho(t)\sigma_- = \begin{pmatrix} \rho_{eg}(t) & 0 \\ \rho_{gg}(t) & 0 \end{pmatrix}, 
\label{eq:rhoppInit}
\end{equation}
being determined by the elements of the ``true'' density matrix $\rho_{eg}(t)=\langle e|\rho(t)|g\rangle$ and $\rho_{gg}(t)=\langle g|\rho(t)|g\rangle$, see Eq.~\ref{eq:TrueDensMatr}. 
Similarly, for $\rho'(t,s)$ the initial condition at $s=t$ are 
\begin{eqnarray}
\rho'_{ee}(t,t)&=&\rho'_{eg}(t,t)=0,\nonumber \\
\rho'_{gg}(t,t)&=&\rho_{eg}(t),\quad \rho'_{ge}(t,t)=\rho_{ee}(t),
\end{eqnarray}
and the corresponding two-time correlator is 
\begin{eqnarray}
\langle\sigma_+(t+\theta)\sigma_-(t)\rangle &=& \mathrm{Tr} \left[ \rho'(t,t+\theta) \sigma_+ \right] \nonumber \\
&=& \rho'_{ge}(t,t+\theta).
 \label{eq:traceRhoPrimeSigmaPlus}
\end{eqnarray}
Therefore, our task is reduced to to determining $\rho''_{ge}(t,t+\theta)$ and $\rho'_{ge}(t,t+\theta)$.

The master equations characterizing the time evolution of the TLS density matrix are given by Eqs.~\ref{eq:MasterEquation_1} and Eq.~\ref{eq:pulseEffect}; the time development of the matrices $\rho''(t,s)$ and $\rho'(t,s)$ also obeys these equations of motion as $s$ increases from $t$ to $t+\theta$. Specifically, when $s$ corresponds to the time interval between the pulses, we have
\begin{eqnarray}
 \frac{d}{ds}\, \rho''_{ee}(t,s) &=& -\Gamma\, \rho''_{ee}(t,s) \\
 \frac{d}{ds}\, \rho''_{gg}(t,s) &=& \Gamma\, \rho''_{ee}(t,s) \\
 \frac{d}{ds}\, \rho''_{ge}(t,s) &=& \left(i\Delta - \frac{\Gamma}{2}\right)\, \rho''_{ge}(t,s) \label{eq:rhoppEOMc}\\
 \frac{d}{ds}\, \rho''_{eg}(t,s) &=& \left(-i\Delta - \frac{\Gamma}{2}\right)\, \rho''_{eg}(t,s) \label{eq:rhoppEOMd}
\end{eqnarray}
for any value of the parameter $t$; the same equations govern the dynamics of $\rho'$. The effect of the pulses on $\rho''$ and $\rho'$ is also easily derived from Eq.~\ref{eq:pulseEffect}: when $t+s$ coincides with the time of the pulse application, i.e.\ when $s=n\tau$ for some integer $n$, the matrix transforms as
\begin{equation}
\rho''(t,n\tau + 0) = \sigma_x \rho''(t,n\tau - 0) \sigma_x 
\end{equation}
where $\rho''(t,n\tau-0)$ and $\rho''(t,n\tau+0)$ are the matrices immediately before and after the pulse, correspondingly; in other words, each pulse interchanges $\rho''_{ee}$ with $\rho''_{gg}$, and $\rho''_{eg}$ with $\rho''_{ge}$; the transformation of $\rho'$ is the same.

\begin{figure}
\includegraphics[width=8.5cm]{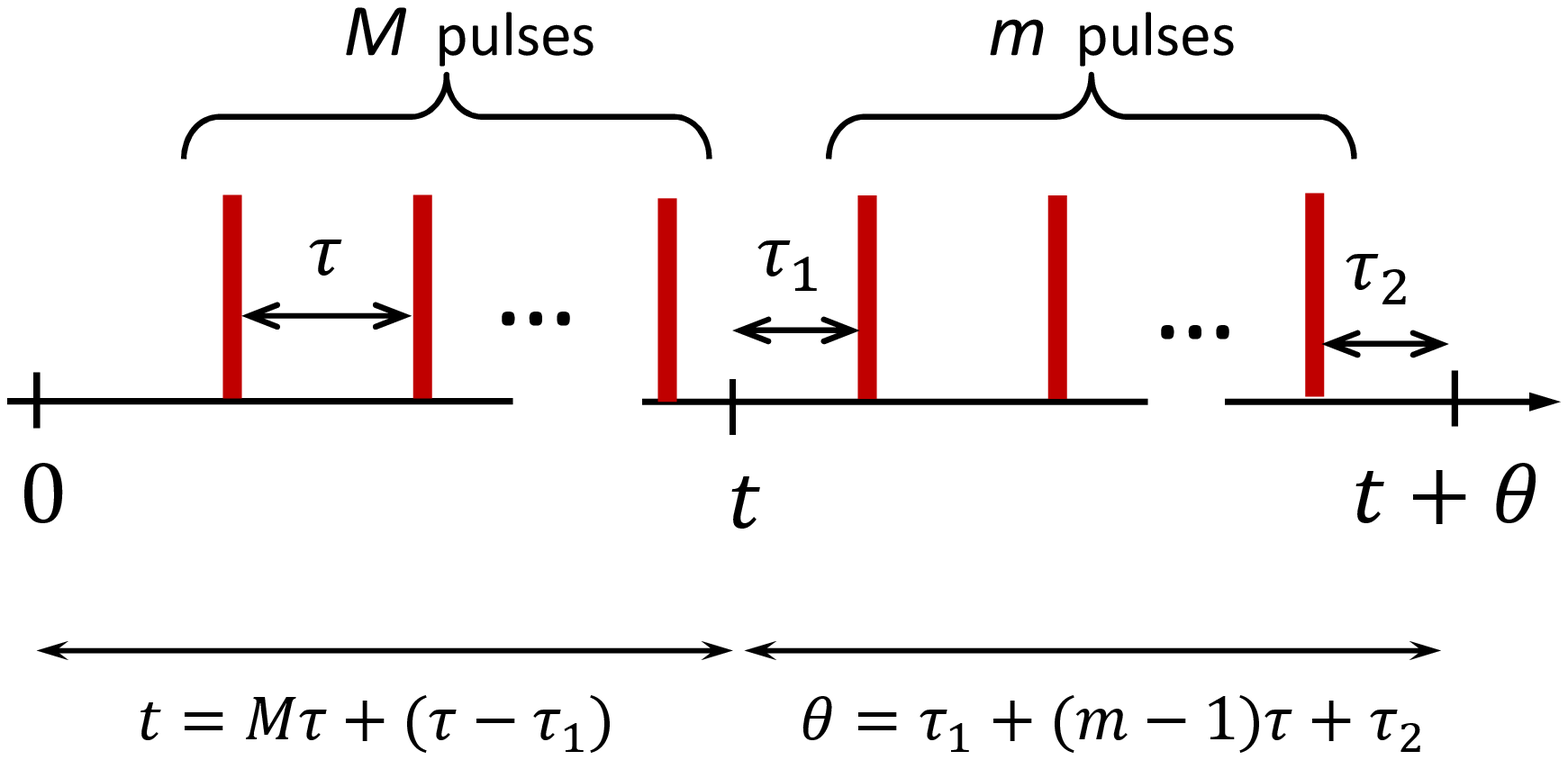} 
\caption{(Color online) Schematic picture of the mutual positions of the time instants $t$ and $t+\theta$ with respect to the pulses.} 
\label{fig:AppendFigPulses}
\end{figure}

Let us start with establishing the initial condition for $\rho''(t,s)$ at $s=t$, which is determined by $\rho_{gg}(t)$ and $\rho_{eg}(t)$, see Eq.~\ref{eq:rhoppInit}. First, we note that $\rho_{eg}(t)\equiv 0$. Indeed, the initial condition at $t=0$ for the density matrix $\rho$ are
\begin{equation}
 \rho_{ee}(0) = 1, \quad  \rho_{gg}(0) = \rho_{ge}(0) = \rho_{eg}(0) = 0.
\end{equation}
As the master equations (\ref{eq:MasterEquation_1}) show, both quantities $\rho_{eg}$ and $\rho_{ge}$ remain zero before the first pulse (when $\Omega_x(t)\equiv 0$). The effect of the pulse is to interchange these two values, i.e.\ they both remain zero after the pulse. The same considerations can be applied for the second, third, etc.\ pulse, showing that $\rho_{eg}(t)=\rho_{ge}(t)=0$ for all $t$. 
Thus, to determine $\rho''(t,t)$ we only need to find $\rho_{gg}(t)$. We assume that the time instant $t$ is between the $M$-th and the $(M+1)$-th pulse, i.e.\ $t = M \tau + (\tau - \tau_1)$ for some $\tau_1\in [0,\tau]$, as shown in Fig.~\ref{fig:AppendFigPulses}. Immediately before the first pulse, at the time moment $\tau-0$, we have:
\begin{eqnarray}
 \rho_{ee}(\tau)&=& \mathrm{e}^{-\Gamma \tau} \nonumber\\
 \rho_{gg}(\tau)&=& 1- \mathrm{e}^{-\Gamma \tau}; 
\end{eqnarray}
then at time $2\tau -0$ we have:
\begin{eqnarray}
 \rho_{ee}(2\tau)&=& (1 - \mathrm{e}^{-\Gamma \tau}) \mathrm{e}^{-\Gamma \tau} \nonumber \\
 \rho_{gg}(2\tau)&=& 1- \mathrm{e}^{-\Gamma \tau} +\mathrm{e}^{-2 \Gamma \tau},
\end{eqnarray}
at time $3\tau -0$: 
\begin{eqnarray}
 \rho_{ee}(3\tau)&=& \mathrm{e}^{-\Gamma \tau}- \mathrm{e}^{-2\Gamma \tau} +\mathrm{e}^{-3 \Gamma \tau} \nonumber \\
 \rho_{gg}(3\tau)&=& 1- \mathrm{e}^{-\Gamma \tau} +\mathrm{e}^{-2 \Gamma \tau}-\mathrm{e}^{-3 \Gamma \tau},
\end{eqnarray}
\begin{center}
 $\vdots$\\
\end{center}
so that eventually, right before the $M$-th pulse, at time $M\tau-0$
\begin{equation}
 \rho_{ee}(M\tau-0) =  \sum_{k=1}^M (-1)^{k-1} \mathrm{e}^{- k\Gamma \tau} = -\sum_{k=1}^M (-1)^{k} \mathrm{e}^{- k\Gamma \tau} ,
\end{equation}
and right after the $M$-th pulse, which interchanges $\rho_{ee}$ and $\rho_{gg}$, 
\begin{equation}
 \rho_{ee}(M\tau+0) =  1- \rho_{ee}(M\tau-0) = 1 + \sum_{k=1}^M (-1)^{k} \mathrm{e}^{- k\Gamma \tau}.
\end{equation}
Thus, at the time instant $t=M\tau+(\tau-\tau_1)$, we have
\begin{eqnarray}
  \rho_{ee}(t)&=& \left( 1+ \sum_{k=1}^M (-1)^{k} \mathrm{e}^{- k\Gamma \tau} \right) \mathrm{e}^{-\Gamma (\tau -\tau_1)} \nonumber\\
	      &=&  \left( \sum_{k=0}^M (-1)^{k} \mathrm{e}^{- k\Gamma \tau} \right) \mathrm{e}^{-\Gamma (\tau -\tau_1)} \nonumber \\
	      &=& \frac{1-(-1)^{M+1} \mathrm{e}^{- (M+1) \Gamma \tau}}{1 + \mathrm{e}^{ - \Gamma \tau}} \mathrm{e}^{-\Gamma (\tau -\tau_1)},
\end{eqnarray}
and, since $\rho_{gg}(t)=1-\rho_{ee}(t)$, we obtain
\begin{equation}
 \rho_{gg}(t) = 1-\frac{1-(-1)^{M+1} \mathrm{e}^{- (M+1) \Gamma \tau}}{1 + \mathrm{e}^{ - \Gamma \tau}} \mathrm{e}^{-\Gamma (\tau -\tau_1)}.
 \label{eq:rho_gg_t}
\end{equation}

Having established the explicit initial value of $\rho''(t,t)$, now we can proceed evaluating the value of $\rho''_{ge}(t,t+\theta)$. Between the pulses both $\rho^{\prime\prime}_{ge}$ and $\rho^{\prime\prime}_{eg}$ evolve according to Eqs.~\ref{eq:rhoppEOMc} and \ref{eq:rhoppEOMd}.
Thus, if $t$ and $t+\theta$ belong to the same inter-pulse interval (i.e.\ when $M\tau\;<t+\theta\;<\;(M+1)\tau$), we have
\begin{equation}
\rho^{\prime\prime}_{ge}(t,t+\theta) = \mathrm{e}^{(i\Delta - \frac{\Gamma}{2})\theta} \rho_{gg}(t) \;\;\; \mathrm{and} \;\;\; \rho^{\prime\prime}_{eg} = 0.
\label{eq:rhoppEvol0}
\end{equation}
With increasing $\theta$, at some point it will become equal to $\tau_1$, and then the instant $t+\theta$ will coincide with the instant when a pulse is applied: $t+\theta=(M+1)\tau+0$. At this point the time instants $t$ and $t+\theta$ will become separated by one pulse, and the value of $\rho''_{eg}$ will be interchanged with $\rho''_{ge}$, i.e.\ when $\theta=\tau_1+0$ we will have already 
\begin{eqnarray}
\rho^{\prime\prime}_{ge}(t,t+\theta) &=& 0 \\
\rho^{\prime\prime}_{eg}(t,t+\theta) &=& \mathrm{e}^{(i\Delta - \frac{\Gamma}{2})\tau_1} \rho_{gg}(t) \nonumber .
\end{eqnarray}
At this point, the accumulation rate of the phase in $\rho''_{ge}$ and $\rho''_{eg}$ changes sign: note that the factors on the right-hand sides of Eqs.~\ref{eq:rhoppEOMc} and \ref{eq:rhoppEOMd} have opposite imaginary parts, $i\Delta$ and $-i\Delta$, respectively. Thus, right before the next pulse, when $t+\theta=(M+2)\tau-0$ (i.e.\ when $\theta=\tau+\tau_1-0$), we will have 
\begin{eqnarray}
\rho^{\prime\prime}_{ge}(t,t+\theta) &=& 0 \\
\rho^{\prime\prime}_{eg}(t,t+\theta) &=& \mathrm{e}^{(i\Delta - \frac{\Gamma}{2})\tau_1} \mathrm{e}^{(-i\Delta - \frac{\Gamma}{2})\tau}\rho_{gg}(t) \nonumber
\end{eqnarray}
and right after the pulse, when $t+\theta=(M+2)\tau + 0$ (i.e.\ when $\theta=\tau+\tau_1 + 0$), the values will be interchanged again:
\begin{eqnarray}
\rho^{\prime\prime}_{ge}(t,t+\theta) &=& \mathrm{e}^{(i\Delta - \frac{\Gamma}{2})\tau_1} \mathrm{e}^{(-i\Delta - \frac{\Gamma}{2})\tau}\rho_{gg}(t) \nonumber\\
\rho^{\prime\prime}_{eg}(t,t+\theta) &=& 0.
\end{eqnarray}
%
Proceeding further in this way, right before the next pulse, at $\theta=\tau_1+2\tau - 0$, we get
\begin{eqnarray}
\rho^{\prime\prime}_{ge}(t,t+\theta) &=& \mathrm{e}^{(i\Delta - \frac{\Gamma}{2})\tau_1} \mathrm{e}^{- \Gamma\tau}\rho_{gg}(t) \nonumber \\
\rho^{\prime\prime}_{eg}(t,t+\theta) &=& 0.
\end{eqnarray}
Note that the phase of $\rho''_{ge}$ still equals to $i\Delta\tau_1$, because after each pulse the phase accumulation rate changes sign. Further, at $\theta=\tau_1+3\tau - 0$, 
\begin{eqnarray}
\rho^{\prime\prime}_{ge}(t,t+\theta) &=& 0 \\
\rho^{\prime\prime}_{eg}(t,t+\theta) &=& \mathrm{e}^{(i\Delta - \frac{\Gamma}{2})\tau_1 -i\Delta \tau - \frac{3\Gamma}{2}\tau}\rho_{gg}(t) \nonumber
\end{eqnarray}
Thus we obtain that for $\theta=\tau_1+(m-1)\tau - 0$ with $m$ even, as shown in Fig.~\ref{fig:AppendFigPulses},
\begin{eqnarray}
\rho^{\prime\prime}_{ge}(t,t+\theta) &=& 0 \\
\rho^{\prime\prime}_{eg}(t,t+\theta) &=& \mathrm{e}^{(i\Delta - \frac{\Gamma}{2})\tau_1 -i\Delta \tau - \frac{(m-1)\Gamma}{2}\tau}\rho_{gg}(t), \nonumber
\end{eqnarray}
and for $\theta=\tau_1+(m-1)\tau + \tau_2$ with $m$ even and $\tau_2 < \tau$,
\begin{eqnarray}
\rho^{\prime\prime}_{ge}(t,t+\theta) &=& \mathrm{e}^{(i\Delta - \frac{\Gamma}{2})\tau_1 -i\Delta \tau - \frac{(m-1)\Gamma}{2}\tau} \mathrm{e}^{i\Delta \tau_2 -\frac{\Gamma}{2} \tau_2 } \rho_{gg}(t) \nonumber \\
		     &=& \mathrm{e}^{-\Gamma \theta /2} \mathrm{e}^{i\Delta ( \tau_1 + \tau_2 - \tau)} \rho_{gg}(t) \nonumber \\
\rho^{\prime\prime}_{eg}(t,t+\theta) &=& 0.
\label{eq:rhoppEvol1}
\end{eqnarray}
Altogether we can write
\begin{equation}
 \rho^{\prime\prime}_{ge}(t,t+\theta) = f(t,\theta) \rho_{gg}(t)
\end{equation}
with $\rho_{gg}(t)$ given above by Eq.~\ref{eq:rho_gg_t}, and
\begin{itemize}
 \item for $t$ and $t+\theta$ in the same pulse interval,
 \begin{equation}
  f(t, \theta) = \mathrm{e}^{(i \Delta - \frac{\Gamma}{2})\theta}
 \end{equation}
\item for $t$ and $t+\theta$ separated by an odd number of pulses,
\begin{equation}
 f(t, \theta) = 0
\end{equation}
\item for $t$ and $t+\theta$ separated by an even number $m$ of pulses, i.e.\ when $\theta=\tau_1+(m-1)\tau+\tau_2$ with even $m$ and $\tau_2<\tau$ (Fig.~\ref{fig:AppendFigPulses}), 
\begin{equation}
 f(t, \theta) = \mathrm{e}^{-\Gamma \theta /2} \mathrm{e}^{i\Delta ( \tau_1 + \tau_2 - \tau)}=\mathrm{e}^{-\Gamma \theta /2} \mathrm{e}^{i\Delta (\theta - m\tau)}.
\end{equation}
\end{itemize}
Note that, due to the pulses, the phase of the function $f(t,\theta)$ does not grow linearly with $\theta$, being confined to the interval $[-\tau\Delta,\tau\Delta]$ at all values of $t$ and $\theta$. This is the reason why, for small inter-pulse delay $\tau\ll\Delta^{-1}$, both emission and absorption are concentrated in the vicinity of $\omega=0$ instead of $\omega=\Delta$.

With this result, we can now rewrite the direct absorption integral $\mathcal{P}_2(\omega)$
in the form
\begin{equation}
 \mathcal{P}_2(\omega) = \int_0^T \; dt \; \rho_{gg}(t) \int_0^{T-t} d\theta f(t, \theta) \mathrm{e}^{-i \omega \theta} 
\end{equation}

First let us evaluate the inner integral, that we will denote as $I_{\theta}$, using the explicit form of $f(t,\theta)$ above:
\begin{widetext}
\begin{eqnarray}
 I_{\theta} &=& \int_0^{\tau_1} \; d\theta \; \mathrm{e}^{-i \omega \theta} \mathrm{e}^{(i\Delta-\Gamma/2)\theta} + \int_{\tau_1+\tau}^{\tau_1 + 2\tau} \; d\theta \; \mathrm{e}^{-i \omega \theta} \mathrm{e}^{-i 2 \Delta \tau + ( i \Delta-\Gamma/2) \theta} \nonumber \\
&+& \int_{\tau_1+3\tau}^{\tau_1 + 4\tau} \; d\theta \; \mathrm{e}^{-i \omega \theta} \mathrm{e}^{-i 4 \Delta \tau + ( i \Delta-\Gamma/2) \theta} + \cdots +
\int_{\tau_1+(m-1)\tau\atop {\rm even}\ m}^{\tau_1 + m\tau} d\theta \; \mathrm{e}^{-i \omega \theta} \mathrm{e}^{-im\Delta \tau + (i \Delta-\Gamma/2) \theta} + \cdots \nonumber \\
&=& \frac{\mathrm{e}^{[i(\Delta -\omega) - \Gamma/2]\tau_1 } -1}{ i(\Delta -\omega) - \Gamma/2} + \sum_{m=2\atop{\rm even}\ m}^{m_{max}} \int_{\tau_1 + (m-1)\tau}^{\tau_1 + m\tau} d\theta \; \mathrm{e}^{-i \omega \theta} \mathrm{e}^{\left[ -i m\Delta \tau + (i \Delta-\Gamma/2) \theta \right] }
\label{eq:Itheta}
\end{eqnarray}
where the summation is over even values of $m$, and $m_{max}$ is the maximum value of $m$; since it has to be even, its specific value depends on whether $M$ is odd or even, see below for details. 
Defining $\gamma_0 = i(\omega -\Delta ) + \Gamma/2$, we can write
\begin{eqnarray}
  I_{\theta} &=& \frac{1 - \mathrm{e}^{-\gamma_0 \tau_1}}{\gamma_0} + \sum_{m=2\atop {\rm even}\ m}^{m_{max}} \mathrm{e}^{ -im\Delta \tau} \frac{\mathrm{e}^{-\gamma_0(\tau_1+(m-1)\tau)}-\mathrm{e}^{-\gamma_0(\tau_1+m\tau)}}{\gamma_0} \nonumber\\
           &=&  \frac{1-\mathrm{e}^{-\gamma_0\tau_1}}{\gamma_0} + \frac{\mathrm{e}^{-\gamma_0\tau_1}}{\gamma_0}(\mathrm{e}^{\gamma_0 \tau}-1) \frac{1-\mathrm{e}^{-m_{max}\gamma_1\tau} }{\mathrm{e}^{2\gamma_1 \tau}-1 },
\end{eqnarray}
\end{widetext}
where we have introduced $\gamma_1 = \Gamma/2 + i\omega$. Note that this result correctly reproduces the situation of $m_{max}<2$, i.e.\ when $m_{max}=0$; this happens when $t$ belongs to the last inter-pulse interval of the sequence, and $\theta$ varies only from zero to $\tau_1$. Then the value of $I_\theta$ is given by the first integral in Eq.~\ref{eq:Itheta}, while the remaining sum over $m$ is zero. Thus, we do not need to worry about this special case in the calculations below. 

Now we need to evaluate the outer integral:
\begin{equation}
\mathcal{P}_2(\omega) =  \int_0^T \; dt \; \rho_{gg}(t) I_{\theta}
\end{equation}
with the quantity $\rho_{gg}$ calculated earlier,
\begin{eqnarray}
 \rho_{gg}(t) &=& 1 - \rho_0(M) \mathrm{e}^{-\Gamma (\tau -\tau_1)} \nonumber\\
              &=& 1 - \frac{ 1- (-1)^{M+1}\mathrm{e}^{-(M+1)\Gamma\tau}}{ 1 + \mathrm{e}^{-\Gamma\tau}}\mathrm{e}^{-\Gamma (\tau -\tau_1)},
\end{eqnarray}
where we introduced the shorthand notation $\rho_0(M)$ for the awkward fraction appearing on the second line. In this way, we represent ${\mathcal P}_2$ as
\begin{eqnarray}
 \mathcal{P}_2 &=& \int_0^T dt \; I_{\theta} - \int_0^{\tau} dt \; \rho_0(0) \mathrm{e}^{-\Gamma(\tau-\tau_1)} I_{\theta} \\
  &-& \int_{\tau}^{2\tau} dt \; \rho_0(1) \mathrm{e}^{-\Gamma(\tau-\tau_1)} I_{\theta} \nonumber \\
  &-& \cdots - \int_{(N_p-1)\tau}^{N_p\tau} dt \; \rho_0(N_p-1) \mathrm{e}^{-\Gamma(\tau-\tau_1)} I_{\theta} \\
&=& \int_0^T dt \; I_{\theta} \nonumber \\
&-& \sum_{M=0}^{N_p-1} \; \rho_0(M) \int_0^{\tau} dt_1 \; \mathrm{e}^{-\Gamma t_1} \left(\frac{1}{\gamma_0} + \mathrm{e}^{-\gamma_0 \tau_1} I_{\theta}^{(1)}\right) \nonumber \\
&&
\end{eqnarray}
where we have defined $t_1 =t-M\tau= \tau - \tau_1$ and
\begin{eqnarray}
I_\theta&=&\frac{1}{\gamma_0} + \mathrm{e}^{-\gamma_0 \tau_1} I_\theta^{(1)}\nonumber\\
 I_{\theta}^{(1)} &=& -\frac{1}{\gamma_0} + \frac{\mathrm{e}^{\gamma_0 \tau} -1}{\gamma_0}\, \frac{1-\mathrm{e}^{-m_{max}\gamma_1\tau}}{\mathrm{e}^{2\gamma_1\tau}-1} \nonumber \\
 &=& -1/\gamma_0 + I_{\theta}^{(2)}
\end{eqnarray}
where we introduced the shorthand notation $I_{\theta}^{(2)}$ for another awkward fraction, the second summand on the second line above.

Now we have
\begin{eqnarray}
 \mathcal{P}_2 = &-& \sum_{M=0}^{N_p-1} \rho_0(M) \int_0^{\tau} dt_1 \; \mathrm{e}^{-\Gamma t_1} \left(\frac{1}{\gamma_0} + \mathrm{e}^{-\gamma_0 (\tau-t_1)} I_{\theta}^{(1)}   \right) \nonumber\\
&+&\int_0^T dt \; I_{\theta} \nonumber \\
= &-& \sum_{M=0}^{N_p-1} \rho_0(M) \left[\frac{1 - \mathrm{e}^{-\Gamma \tau}}{\gamma_0 \Gamma}  + I_{\theta}^{(1)} \mathrm{e}^{-\gamma_0 \tau}  \frac{\mathrm{e}^{\gamma_2 \tau} -1}{\gamma_2}  \right] \nonumber\\
&+& \mathcal{P}_3
\label{eq:P2withP3}
\end{eqnarray}
with $ \mathcal{P}_3 = \int_0^T dt \; I_{\theta}$ and $\gamma_2 = \gamma_0 - \Gamma = i(\omega - \Delta) - \frac{\Gamma}{2}$. 

Below we will show that the first sum gives exactly the contribution from the stimulated emission $\mathcal{P}_1(\omega)$. It is convenient to calculate the simpler term $\mathcal{P}_3$ first. We can rewrite $\mathcal{P}_3$ as:
\begin{eqnarray}
 \mathcal{P}_3 &=& \int_0^T \frac{dt}{\gamma_0} +  \int_0^{\tau} dt \; \mathrm{e}^{-\gamma_0 \tau_1} I_{\theta}^{(1)} \nonumber \\
&+& \cdots + \int_{(N_p-1)\tau}^{N_p\tau} dt \; \mathrm{e}^{-\gamma_0 \tau_1} I_{\theta}^{(1)} \nonumber \\
&=& \frac{N_p\tau}{\gamma_0} + \mathrm{e}^{-\gamma_0 \tau} \sum_{M=0}^{N_p-1} I_{\theta}^{(1)} \int_0^{\tau} dt_1 \; \mathrm{e}^{\gamma_0 t_1} \nonumber \\
&=& \frac{N_p\tau}{\gamma_0} + \sum_{M=0}^{N_p-1} I_{\theta}^{(1)} \frac{1- \mathrm{e}^{-\gamma_0 \tau} }{ \gamma_0} 
\end{eqnarray}
%
With the explicit form of $I_\theta^{(1)}$ given above, we have 
\begin{eqnarray}
\mathcal{P}_3 
&=& \frac{N_p\tau}{\gamma_0} - \sum_{M=0}^{N_p-1} \frac{1 - \mathrm{e}^{-\gamma_0 \tau }}{\gamma_0^2} + \sum_{M=0}^{N_p-1} \frac{1 - \mathrm{e}^{-\gamma_0 \tau }}{\gamma_0^2} I_{\theta}^{(2)} \nonumber \\
&=& \frac{N_p\tau}{\gamma_0} - \frac{N_p}{\gamma_0^2}(1 - \mathrm{e}^{-\gamma_0 \tau }) + \mathcal{P}_4
\end{eqnarray}
where 
\begin{eqnarray}
 \mathcal{P}_4 &=& \frac{1-\mathrm{e}^{-\gamma_0\tau}}{\gamma_0}\, \frac{\mathrm{e}^{\gamma_0\tau}-1}{\gamma_0} \sum_{M=0}^{N_p-1} \frac{1-\mathrm{e}^{-\gamma_1\tau m_{max}}}{\mathrm{e}^{2\gamma_1 \tau}-1} \nonumber \\
%
&=& \frac{\mathrm{e}^{\gamma_0 \tau}+\mathrm{e}^{-\gamma_0 \tau}-2}{\gamma_0^2 (\mathrm{e}^{2\gamma_1\tau}-1)} \left[ N_p -\sum_{M=0}^{N_p-1} \mathrm{e}^{-\gamma_1\tau m_{max}} \right]\nonumber \\
\label{eq:P4}
\end{eqnarray}

In order to calculate the last sum in the equation above, we need to determine $m_{max}$. To do this let us consider the case of $N_p=2K$, i.e.\ when the number $K$ of the full cycles of the sequence has been applied to the TLS. Let us recall that we represent $t=M\tau+(\tau-\tau_1)$, i.e.\ $M$ is the number of pulses between zero and $t$. The number of pulses separating $t$ and $t+\theta$ is $m$, and the maximum value of $\theta$ is $\theta_{max}=T-t$, which limits the maximum value of $m$; however, $f(t,\theta)$ is zero if $m$ is odd, so that $m_{max}$ should be even.
Therefore, starting from larger values of $t$, we obtain:
\begin{itemize}
 \item[*] if $t\in[T-\tau,T]$ then $\theta_{max}=(T-t) \in[0,\tau]$, so that if $M=2K-1$ then $m_{max}=0$,
 \item[*] if $t\in[T-2\tau,T-\tau]$ then $\theta_{max}=(T-t) \in[\tau,2\tau]$, so that if $M=2K-2$ then $m_{max}=0$ because $m_{max}$ should be even,
 \item[*] if $t\in[T-3\tau,T-2\tau]$ then $\theta_{max}=(T-t) \in[2\tau,3\tau]$, so that if $M=2K-3$ then $m_{max}=2$,
 \item[*] if $t\in[T-4\tau,T-3\tau]$ then $\theta_{max}=(T-t) \in[3\tau,4\tau]$, so that if $M=2K-4$ then $m_{max}=2$ (should be even),
 \item[*] if $t\in[T-5\tau,T-4\tau]$ then $\theta_{max}=(T-t) \in[4\tau,5\tau]$, so that if $M=2K-5$ then $m_{max}=4$, 
 \item[*] if $t\in[T-6\tau,T-5\tau]$ then $\theta_{max}=(T-t) \in[5\tau,6\tau]$, so that if $M=2K-6$ then $m_{max}=4$ (should be even),
\item[*] $\cdots$
 \item[*] if $t\in[\tau,2\tau]$ then $\theta_{max}=(T-t) \in[T-2\tau,T-\tau]$, so that if $M=1$ then $m_{max}=2K-2$,
 \item[*] if $t\in[0,\tau]$ then $\theta_{max}=(T-t) \in[T-\tau,T]$, so that if $M=0$ then $m_{max}=2K-2$ (should be even).
\end{itemize}
To summarize, if we parametrize $M=2n$ for even $M$ and $M=2n+1$ for odd $M$, where $n$ varies from 0 to $K-1$, then $m_{max}=2(K-n-1)$ for both $M=2n$ and $M=2n+1$.

Thus, the last sum in Eq.~\ref{eq:P4} is calculated as
\begin{eqnarray}
\sum_{M=0}^{N_p-1} \mathrm{e}^{-\gamma_1\tau m_{max}} &=& 2\sum_{n=0}^{K-1} \mathrm{e}^{-2\gamma_1\tau (K-n-1)} \nonumber\\
&=& 2\frac{1-\mathrm{e}^{-\gamma_1\tau N_p}}{1-\mathrm{e}^{-2\gamma_1\tau}},
\end{eqnarray}
where the factor 2 appears because $m_{max}$ is the same for both $M=2n$ and $M=2n+1$, so that the sums over odd $M$ and even $M$ are combined. Putting all terms together, we obtain
\begin{widetext}
\begin{equation}
 \mathcal{P}_3 =  \frac{N_p\tau}{\gamma_0} -  \frac{N_p}{\gamma_0^2}(1 - \mathrm{e}^{-\gamma_0 \tau })
+ \frac{\mathrm{e}^{\gamma_0 \tau}+\mathrm{e}^{-\gamma_0 \tau}-2}{\gamma_0^2(\mathrm{e}^{2\gamma_1\tau}-1)}\, \left[N_p - 2\frac{1-\mathrm{e}^{-\gamma_1\tau N_p}}{1-\mathrm{e}^{-2\gamma_1\tau}}\right]
\end{equation}
\end{widetext}

Now, the calculation of the emission term $\mathcal{P}_1$ can be simplified if we notice that $\rho'(t,s)$ obeys the same equations of motion as $\rho''(t,s)$, and is transformed by the pulses in exactly the same way. Therefore, the quantity $\rho'_{ge}(t,s)$ (that determines $\mathcal{P}_1$) evolves in exactly the same way as $\rho''_{ge}(t,s)$, and the difference between them is only in the initial condition: at $s=t$ we have $\rho'_{ge}(t,t)=\rho_{ee}(t)$, while $\rho''_{ge}(t,t)=\rho_{gg}(t)=1-\rho_{ee}(t)$. Thus, the reasoning that was used in deriving 
Eqs.~\ref{eq:rhoppEvol0}--\ref{eq:rhoppEvol1} can be directly applied to $\rho'_{ge}(t,s)$ if $\rho_{gg}(t)$ is substituted by $\rho_{ee}(t)$, due to the linearity of the master equations. As a result, we immediately see that $\rho'_{ge}(t,t+\theta)$ has the form 
\begin{equation}
 \rho'_{ge}(t,t+\theta) = f(t,\theta) \rho_{ee}(t)
\end{equation}
with the same function $f(t,\theta)$. Thus, the integral $I_\theta$ can be used without modifications in the calculation of $\mathcal{P}_1$, and, since $\rho_{gg}(t)=1-\rho_{ee}(t)$, we immediately obtain
\begin{equation}
 \mathcal{P}_1(\omega) = \int_0^T \; \rho_{ee}(t) I_\theta\; dt  = \int_0^T \; [1-\rho_{gg}(t)] I_\theta\; dt = \mathcal{P}_3 - \mathcal{P}_2.
\end{equation}
Comparing this expression with Eq.~\ref{eq:P2withP3} above, we obtain an explicit expression
\begin{eqnarray}
 \mathcal{P}_1 &=& \sum_{M=0}^{N_p-1} \rho_0(M) \int_0^{\tau} dt_1 \; \mathrm{e}^{-\Gamma t_1} \left(\frac{1}{\gamma_0} + \mathrm{e}^{-\gamma_0 (\tau-t_1)} I_{\theta}^{(1)}   \right) \nonumber\\
&=& \sum_{M=0}^{N_p-1} \rho_0(M) \left[\frac{1 - \mathrm{e}^{-\Gamma \tau}}{\gamma_0 \Gamma}  + I_{\theta}^{(1)} \mathrm{e}^{-\gamma_0 \tau}  \frac{\mathrm{e}^{\gamma_2 \tau} -1}{\gamma_2}  \right]
\label{eq:P1}
\end{eqnarray}

\begin{figure}[htbp]
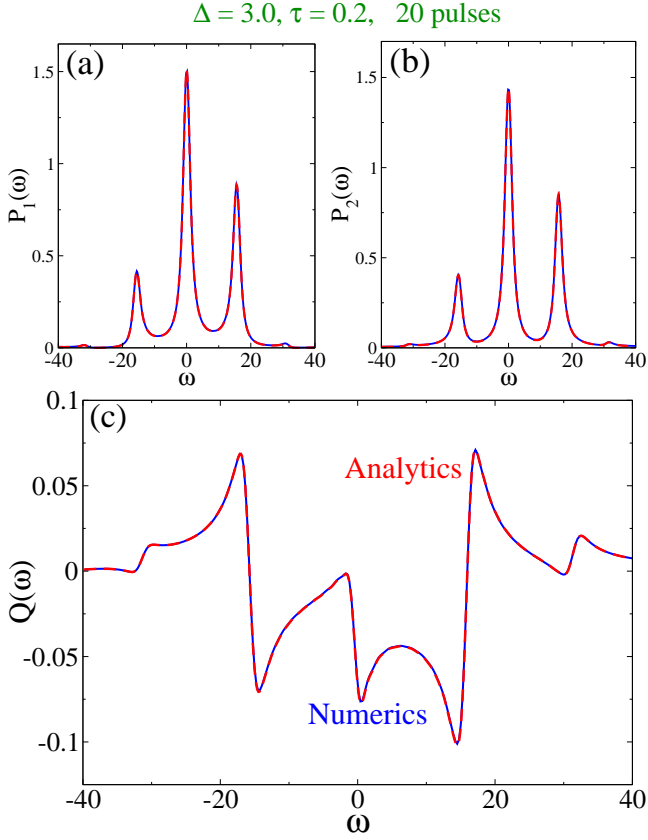

\includegraphics*[width=8.5cm]{figure6_a_b.eps}
\includegraphics*[width=8.5cm]{figure6_c.eps} 
\caption{(Color online) Absorption spectrum of a two-level system with detuning $\Delta = 3.0$ driven by a periodic sequence of $\pi$-pulses of period $\tau = 0.2$ after $20$ pulses. The results are obtained by solving master equation numerically (blue) or analytically in the limit of a large number of pulses (red dashed). We see that indeed the agreement between the numerical and analytical results is better at longer times (for a large number of pulses).}
\label{fig:numerics_analytics_largeNP}
\end{figure}

Now let us evaluate the sums appearing in this expression. First, we need the sum
\begin{eqnarray}
\sum_{M=0}^{N_p-1} &\rho_0(M)& = \frac{1}{1 + \mathrm{e}^{-\Gamma \tau}} \sum_{M=0}^{N_p-1} \left[ 1+\mathrm{e}^{-\Gamma \tau} (-\mathrm{e}^{-\Gamma \tau})^M   \right] \nonumber\\
   &=& \frac{N_p}{1+\mathrm{e}^{-\Gamma\tau}} + \frac{\mathrm{e}^{-\Gamma \tau}}{1+\mathrm{e}^{-\Gamma \tau}} \frac{1 -(-\mathrm{e}^{-\Gamma \tau})^{N_p}}{1 + \mathrm{e}^{-\Gamma \tau}}
\end{eqnarray}
and for sufficiently large $N_p$, when the exponentially small terms can be omitted, this yields
\begin{equation}
\sum_{M=0}^{N_p-1} \rho_0(M) \approx \frac{N_p}{1+ \mathrm{e}^{-\Gamma \tau}} + \frac{\mathrm{e}^{-\Gamma \tau}}{(1 + \mathrm{e}^{-\Gamma \tau})^2},
\end{equation}
The second required sum is
\begin{equation}
\sum_{M=0}^{N_p-1} \rho_0(M) \mathrm{e}^{-\gamma_1 \tau m_{max}},
\end{equation}
and in order to evaluate it we use the same parametrization as above, $M=2n$ for even $M$ and $M=2n+1$ for odd $M$, with $n=0,\dots,K-1$. We pair the neighboring terms, i.e.\
\begin{eqnarray}
 \sum_{M=0}^{N_p-1} &\rho_0&(M) \mathrm{e}^{-\gamma_1 \tau m_{max}} \\
&= & \left[\sum_{{\rm even} M} +\sum_{{\rm odd} M}\right] \rho_0(M) \mathrm{e}^{-\gamma_1 \tau m_{max}} \nonumber\\
&= &  \sum_{n=0}^{K-1} \left[ \rho_0(2n) + \rho_0(2n+1)\right] \mathrm{e}^{-2\gamma_1 \tau (K-n-1)}. \nonumber
\end{eqnarray}
Since 
\begin{equation}
 \rho_0(2n) + \rho_0(2n+1) = \frac{2 + \mathrm{e}^{-\Gamma \tau (2n+1)} - \mathrm{e}^{-\Gamma \tau (2n+2) }}{ 1 + \mathrm{e}^{-\Gamma \tau}},
\end{equation}
we obtain
\begin{eqnarray}
\sum_{M=0}^{N_p-1} &\rho_0&(M) \mathrm{e}^{-\gamma_1 \tau m_{max}}\nonumber \\
&=& \sum_{n=0}^{K-1} \frac{2 - \mathrm{e}^{-\Gamma \tau (2n+1)} + \mathrm{e}^{-\Gamma \tau (2n+2) }}{ 1 + \mathrm{e}^{-\Gamma \tau}} \mathrm{e}^{-2\gamma_1 (K-n-1) \tau} \nonumber \\
&=& \frac{\mathrm{e}^{-2\gamma_1 (K-1)\tau}}{ 1 + \mathrm{e}^{-\Gamma \tau}} \left[2\frac{\mathrm{e}^{2K\gamma_1\tau}-1}{\mathrm{e}^{2\gamma_1\tau}-1} \right. \nonumber \\
&+& \left. \mathrm{e}^{-\Gamma \tau}(1- \mathrm{e}^{-\Gamma \tau}) \frac{\mathrm{e}^{2K(\gamma_1-\Gamma)\tau}-1}{\mathrm{e}^{2(\gamma_1-\Gamma)\tau}-1} \right]  
\label{eq:sumrho0}
\end{eqnarray}
%



Substituting these results into Eq.~\ref{eq:P1} for $\mathcal{P}_1$, we obtain in the limit of large $N_p$

\begin{widetext}
\begin{eqnarray}
 \mathcal{P}_1(\omega) &=& \frac{1}{(1 + \mathrm{e}^{-\Gamma \tau})\gamma_0} \Bigg[ \left( \frac{1-\mathrm{e}^{-\Gamma \tau}}{\Gamma} - \mathrm{e}^{-\gamma_0 \tau} \frac{\mathrm{e}^{\gamma_2 \tau}-1}{\gamma_2}
 + \frac{\mathrm{e}^{\gamma_2 \tau}-1}{\gamma_2}\,\frac{1-\mathrm{e}^{-\gamma_0 \tau}}{\mathrm{e}^{2 \gamma_1 \tau}-1} \right) 
\left(N_p + \frac{\mathrm{e}^{-\Gamma \tau}}{1 + \mathrm{e}^{-\Gamma \tau}} \right) \nonumber \\
&-& \frac{\mathrm{e}^{\gamma_2 \tau}-1}{\gamma_2}\, \frac{1-\mathrm{e}^{-\gamma_0 \tau}}{\mathrm{e}^{2 \gamma_1 \tau}-1} \, \left(  
2\,\frac{\mathrm{e}^{-N_p \gamma_1 \tau} - 1 }{ \mathrm{e}^{-2 \gamma_1 \tau } - 1} + (\mathrm{e}^{-\Gamma \tau} - \mathrm{e}^{ - 2 \Gamma \tau})\frac{ \mathrm{e}^{-N_p \gamma_1 \tau } }{\mathrm{e}^{-2 \gamma_1 \tau} - \mathrm{e}^{-2 \Gamma \tau } } \right) 
\Bigg]
\end{eqnarray}
\end{widetext}
and the net absorption spectrum is obtained as
\begin{eqnarray}
  Q(\omega) &=& 2 A^2 \mathrm{Re} \left\{  \mathcal{P}_2(\omega) - \mathcal{P}_1(\omega) \right\} \nonumber\\
	&=& 2 A^2 \mathrm{Re} \left\{ \mathcal{P}_3(\omega) -2\mathcal{P}_1(\omega) \right\}
\end{eqnarray}
with the explicit analytical expressions for $\mathcal{P}_1$ and $\mathcal{P}_3$ given above.

In Fig.\ref{fig:numerics_analytics_largeNP} we show a comparison of the numerical result and the analytical result described above for the absorption spectrum of a two-level system with detuning $\Delta = 3.0$ driven by a periodic pulse sequence of period $\tau = 0.2$ after $20$ pulses. The comparison reveals a very good agreement between the solutions, and the agreement indeed improves as the number of pulses increases.

\end{document}